      [1999/12/01 v1.4c Il Nuovo Cimento]
\documentclass{cimento}
\usepackage{graphicx}  
\title{The updated $E_{\rm peak}$--$E_{\gamma}$ correlation in GRBs}
\author{G. Ghirlanda\from{ins:1}, G. Ghisellini\from{ins:1},
D. Lazzati\from{ins:2}, 
C. Firmani\from{ins:1}}
\instlist{\inst{ins:1} INAF - Osservatorio Astronomico Brera 
  Via E. Bianchi 46, Merate, IT \inst{ins:2} JILA, 440 UCB, University of 
Colorado, Boulder}
\PACSes{\PACSit{98.70.Rz}{Gamma Ray Bursts.}}
\begin{document}

\maketitle

\begin{abstract}
The recently discovered correlation between the rest frame GRB peak
spectral energy $E_{\rm{peak}}$ and the collimation corrected energy
$E_\gamma$ in long GRBs is potentially very important, yet awaits
confirmation from an independent sample. It may help to shed light on
the radiation mechanism of the prompt GRB phase and on the way -- and
in which form -- the energy is released from the central engine. We
here present some additional evidence for the correlation (two new
bursts) and re-derive the best-fit parameters. The tightness of the
correlation is confirmed (sigma=0.1 dex). We show that this
correlation allows us, for the first time, to use GRBs as cosmological
probes to constrain the expansion history of the universe.
\end{abstract}

\section{Introduction}

Since their discovery, GRBs turned out to be incredibly powerful
sources, with detected fluences up to $>10^{-4}$ erg/cm$^2$ in the
$\gamma$--ray band, above few tens of keV.  The first spectroscopic
measurements of their redshifts (e.g.\cite{ref:djo}), besides
confirming their cosmological nature, indicated that these events
release, in the $\gamma$--ray band, a huge amount of energy, up to
$E_{\rm iso}=10^{55}$ erg.  This extraordinary energetic content
became a challenge for the proposed GRB models.  One implicit
hypothesis in deriving the GRB energy, from the observed fluence and
measured redshift, consisted in assuming that GRBs emit isotropically.
However, it was suggested~\cite{ref:sar} that GRBs might be collimated
into a cone of semiaperture $\theta_{\rm j}$.  The jet opening angle
could be directly estimated, under some simplifying assumptions on few
other parameters, from the measure of the achromatic break time $t_b$
in the afterglow light curve \cite{ref:roa}.  The presence of a jet in
GRB outflows, supported by observations in most events \cite{ref:pan},
allowed to correct the isotropic equivalent energy $E_{\rm iso}$ for
the collimation factor, therefore obtaining the collimation corrected
energy $E_{\gamma}=E_{\rm iso}(1-\cos\theta_{\rm j})$
\cite{ref:fra,ref:blo}.  These results suggested that GRBs
might have a unique energy $\sim 10^{51}$ erg.
The study of the rest frame spectral properties of a sample of
$Beppo$SAX GRBs \cite{ref:ama} (see also \cite{ref:llo}) led to the
discovery of a correlation between the isotropic equivalent energy
$E_{\rm iso}$ and the $\nu F_{\nu}$ spectral peak energy, $E_{\rm peak}$ 
(the Amati correlation).

With  the largest  sample  of bursts  with spectroscopically  measured
redshifts, published  spectra and well  determined jet break  time, we
estimated  the  jet  opening  angles  and  derived  $E_{\gamma}$.   We
discovered  a  very tight  correlation  between  $E_{\gamma}$ and  the
spectral  peak  energy $E_{\rm  peak}$  \cite{ref:ghi} (the  Ghirlanda
correlation).  This correlation relates the GRB prompt emission energy
-- properly  corrected   for  the  burst  geometry  --   to  its  peak
frequency.  It might  be  the  key to  understand  some still  obscure
aspects of the physics and  origin of GRBs. Besides, its small scatter
and good  powerlaw fit  allowed to  use, for the  first time,  GRBs as
standard  candles  to   constrain  the  cosmological  parameters  (see
\cite{ref:ghia,ref:fir,ref:ghis}).

\section{The $E_{\rm peak}-E_{\gamma}$ Correlation}

We collected all the GRBs with redshift measurements and published
spectral parameters.  In most cases the spectrum is represented by the
empirical Band function.  The spectrum allows to compute the burst
bolometric fluence $F$ (i.e.  the time integrated flux) and, hence,
the isotropic equivalent energy , i.e.  
$E_{\rm iso}=4\pi D^{2}_{L}(z) F/(1+z)$.
The jet break  time, typically observed between 0.1  and 10 days since
the burst trigger, is due to the deceleration, by the external ISM, of
the GRB relativistic fireball. When the Lorentz factor of the fireball
is $\Gamma\propto 1/\theta_{\rm  j}$ a change in the  time decay slope
of the  afterglow flux is observed.  This  characteristic time depends
also on the  ISM density $n$ and on the kinetic  energy which is still
in the fireball after  the prompt emission phase (parametrized through
the  efficiency $\eta_{\gamma}$).  Therefore,  the measure  of $t_{b}$
allows   to  estimate   $\theta_{\rm  j}\propto   t_{b}^{3/8}  E_{\rm
iso}^{-1/8} (\eta_{\gamma} n)^{1/8}$.  
\begin{figure}
\hspace{-0.2 cm}
\vspace{-0.3 cm}
\resizebox{12.6cm}{10.5cm}
{\includegraphics{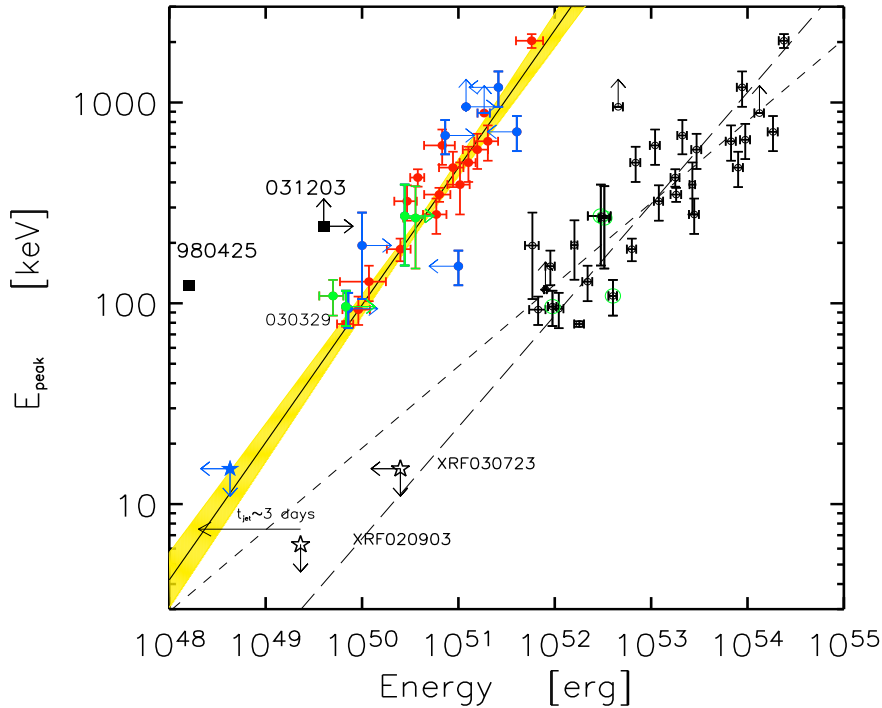}}     
\caption{
The  rest frame $E_{\rm  peak}$--$E_{\rm iso}/E_\gamma$  plane.  Black
open symbols  represent the  isotropic equivalent energy.   Red filled
symbols are  the 15 GRBs for which  a jet break was  measured in their
afterglow light curves (from Tab. 1 and 2 of GGL04).  Blue symbols are
upper/lower limits  for $E_\gamma$. The four new  GRBs are represented
as open green  circles for $E_{\rm iso}$ and  filled green symbols for
$E_{\gamma}$. Also  shown are two outliers (black  squares) for either
the Amati  and the Ghirlanda correlation (filled  squares).  Stars are
the two XRF with known redshift.The Amati correlation is also reported
either fitting with the errors  on both coordinates (long dashed line)
and with the least square  method (dashed line). The best fit Ghirlanda
correlation (solid  black line), giving a reduced  $\chi^2=1.33$ and a
slope  $\sim0.7$, is also  shown with  its uncertainty  region (shaded
area). }
\end{figure}
After the publication of the original work of GGL04, the redshifts and
spectral parameters of four more bursts were published.  We present
here the updated correlations (either $E_{\rm peak}$--$E_{\rm iso}$
and $E_{\rm peak}$--$E_{\gamma}$).  A continuously updated version of
the correlations and the relative tables can be found at
http://www.merate.mi.astro.it/$\sim$ghirla/deep/blink.htm (with the
complete reference list).  We use $\Omega_{\rm M}=0.3$,
$\Omega_\Lambda = h_0=0.7$.  The GRBs added to the original sample of
23 (see also \cite{ref:ghib}, \cite{ref:ghiba} ) are:
\begin{itemize}
\item GRB~021004 \cite{ref:mir},  with a  
rest  frame  peak  energy  $E_{\rm peak}=266\pm116$  keV  and  $E_{\rm
iso}=3.27\pm0.39 \times  10^{52}$ erg.  The jet break  time is $t_{\rm
b}=4.74\pm0.5$   days,   the   jet  opening   angle   $\theta_{\rm
j}=8.51^\circ\pm1.04^\circ$             and,            therefore,
$E_{\gamma}=3.6\pm1.0\times10^{50}$ erg.
\vspace{-0.2cm}
\item GRB~030323 \cite{ref:att},  
with a rest frame peak energy $E_{\rm peak}=272\pm188$ keV and $E_{\rm
iso}=3.0\pm0.8 \times 10^{52}$ erg.  The afterglow light curve is
relatively flat i.e. indicating a jet break time $t_{\rm b}>4.8$ days,
implying $\theta_{\rm j}>7.8^\circ$ and, therefore,
$E_{\gamma}>2.77\times10^{50}$ erg.
\vspace{-0.2cm}
\item GRB~040924 \cite{ref:wie}, with a rest frame peak energy of $E_{\rm peak}=96\pm 20$ 
keV  and $E_{\rm  iso}=9.5\pm1.0 \times  10^{51}$ erg.   The afterglow
light curve  is relatively  flat [$F(t)\propto t^{-1}$]  up to  1 day,
indicating a  jet break time $t_{\rm b}>1$  day, implying $\theta_{\rm
j}>6.9^\circ$ and, therefore, $E_{\gamma}>6.8\times10^{49}$ erg.
\vspace{-0.2cm}
\item GRB~041006 \cite{ref:pri}, 
with a rest frame peak energy $E_{\rm peak}=109\pm22$ keV and $E_{\rm
iso}=4.0\pm0.4 \times 10^{52}$ erg.  The jet break time is $t_{\rm
b}=0.14\pm0.02$ days, implying $\theta_{\rm j}=2.9^\circ\pm0.4^\circ$
and, therefore, $E_{\gamma}=4.9\pm1.3\times10^{49}$ erg.
\end{itemize}
We added the four GRBs to the 23 GRBs of Tab.1 and 2 of GGL04. In Fig.
1 we report the updated correlations.  With 27 GRBs (black symbols is
Fig.  1) the best fit powerlaw (weighting for the errors on both
coordinates) of the $E_{\rm peak}$--$E_{\rm iso}$ correlation is:
\begin{equation}
E_{\rm  peak}/100  {\rm  keV}  \, =\,  
(3.21\pm0.11) (E_{\rm iso}/1.1\times  10^{53}  {\rm
erg})^{0.56\pm0.02}
\end{equation}
with a reduced $\chi^{2}=5.19$ (long dashed line in Fig.  1).  The
least square fit to the same data points (i.e. ignoring the errors on
the two coordinates) gives a slope of 0.41$\pm$0.05 (dashed line in
Fig. 1).

The Ghirlanda  $E_{\rm peak}$--$E_{\gamma}$ correlation,  updated with
two of the four GRBs with known $\theta_{\rm j}$ (green filled symbols
in Fig. 1), is:
\begin{equation}
E_{\rm  peak}/100  {\rm  keV} \, =\, 
(2.5\pm1.0) (E_{\gamma}/3.8\times  10^{50}  {\rm
erg})^{0.69\pm0.04}
\end{equation}
with a reduced $\chi^{2}=1.33$ (solid line in Fig. 1, shaded region is
the uncertainty of this correlation). Least square fit gives a
slightly flatter slope of 0.6.  The present gaussian fit of the
distribution of the perpendicular scatter of the 27 GRB around their
best fit (Eq.  1) is $\sigma=0.22$.  The scatter of the Ghirlanda
correlation (Eq.  2), instead, is only $\sigma=0.1$ (i.e.  consistent
with what found with the 15 GRBs in GGL04).
We stress that the Ghirlanda correlation is well fitted (also with the
2 new GRBs) by a single powerlaw and that its slope, 0.69, is
consistent with what found with the 15 GRBs in GGL04. Moreover, the
reduced $\chi^2=1.33$ allows its use in cosmology (GGLF04, GGF05 and
\cite{ref:ghis}). This is  in net contrast with what claimed by
\cite{ref:fri}  who find  $\chi^2\sim3.71$  for the  
$E_{\rm{peak}}$--$E_{\gamma}$ fit.  Although the authors do not
investigate the reason of their statistically unacceptable results, we
note that their severe underestimation of the uncertainty on two
relevant parameters (i.e.  the ISM density and the jet break time) is
driving their conclusions.
They assume, when unknown, an ISM density with an uncertainty of 50\%.
This parameter is highly uncertain due to the few measured values (and
also in these few cases highly debated).  Until precise measurements
of this parameter will not be available, it is preferable to let it
vary within a relatively large range.  This is indeed what GGL04 did
in their original work (where $1<n<10$).  Moreover, in \cite{ref:fri}
a few jet break times are reported with unreliably small errors (down
to 1\%).  This means that for some GRBs we could determine a break in
their light curves at -- say -- 1.5 days with an uncertainty of 28
minutes (e.g. GRB~011211), which may be a challenge for the future but
which was extremely hard in the past afterglow observational
campaigns.  

\section{Conclusions}

We have presented the updated Ghirlanda correlation with 17 GRBs with
firm redshift measurements and published spectral parameters. These 2
more events perfectly fit the Ghirlanda correlation as found by
GGL04. While waiting for future events, even the present small sample
of GRBs have important and intriguing implications for the use of GRBs
as standard candles to measure our universe.
\vspace{-0.1cm}
\acknowledgments
We thank Annalisa Celotti \& Fabrizio Tavecchio for useful
discussions.  GG and GG thank the MIUR for the Cofin grant
2003020775\_002.

\end{document}